# Security Risks in Deep Learning Implementations

Qixue Xiao[1], Kang Li[2], Deyue Zhang[1], Weilin Xu[3]

[1]Qihoo 360 Security Research Lab   [2]University of Georgia   [3]University of Virginia

*Abstract*—Advance in deep learning algorithms overshadows their security risk in software implementations. This paper discloses a set of vulnerabilities in popular deep learning frameworks including Caffe, TensorFlow, and Torch. Contrast to the small code size of deep learning models, these deep learning frameworks are complex and contain heavy dependencies on numerous open source packages. This paper considers the risks caused by these vulnerabilities by studying their impact on common deep learning applications such as voice recognition and image classifications. By exploiting these framework implementations, attackers can launch denial-of-service attacks that crash or hang a deep learning application, or control-flow hijacking attacks that cause either system compromise or recognition evasions. The goal of this paper is to draw attention on the software implementations and call for the community effort to improve the security of deep learning frameworks.

## I. INTRODUCTION

Artificial intelligence becomes an attention focus in recent years partially due to the success of deep learning applications. Advances in GPUs and deep learning algorithms along with large datasets allow deep learning algorithms to address real-world problems in many areas, from image classification to health care prediction, and from auto game playing to reverse engineering. Many scientific and engineering fields are passionately embracing deep learning.

These passionate adoptions of new machine learning algorithms has sparked the development of multiple deep learning frameworks, such as Caffe [3], TensorFlow [1], and Torch [6]. These frameworks enable fast development of deep learning applications. A framework provides common building blocks for layers of a neural network. By using these frameworks, developers can focus on model design and application specific logic without worrying the coding details of input parsing, matrix multiplication, or GPU optimizations.

In this paper, we examine the implementation of three popular deep learning frameworks: Caffe, TensorFlow, and Torch. And we collected their software dependencies based on the sample applications released along with the framework. The implementation of these frameworks are complex (often with hundreds of thousands lines of code) and are often built over numerous 3rd party software packages, such as image and video processing, scientific computation libraries.

A common challenge for the software industry is that implementation complexity often leads to software vulnerabilities. Deep learning frameworks face the same challenge. Through our examination, we found multiple dozens of implementation flaws. Among them, 15 ones have been assigned with CVE numbers. The types of flaws cover multiple common types of software bugs, such as heap overflow, integer overflow, use-after-free.

We made a preliminary study on the threats and risks caused by these vulnerabilities. With a wide variety of deep learning applications being built over these frameworks, we consider a range of attack surfaces including malformed data in application inputs, training data, and models. The potential consequences from these vulnerabilities include denial-of-service attack, evasion to classifications, and even to system compromise. This paper provides a brief summary of these vulnerabilities and the potential risks that we anticipate for deep learning applications built over these frameworks.

Through our preliminary study of three deep learning frameworks, we make the following contributions:

- This paper exposes the dependency complexity of popular deep learning frameworks.
- This paper presents a preliminary study of the attack surface for deep learning applications.
- Through this paper, we show that multiple vulnerabilities exist in the implementation of these frameworks.
- We also study the impact of these vulnerabilities and describe the potential security risks to applications built on these vulnerable frameworks.

## II. LAYERED IMPLEMENTATION OF DEEP LEARNING APPLICATIONS

Deep learning frameworks enable fast development of machine learning applications. Equipped with pre-implemented neural network layers, deep learning frameworks allow developers to focus on application logic. Developers can design, build, and train scenario specific models on a deep learning framework without worrying about the coding details of input parsing, matrix multiplication, or GPU optimizations.

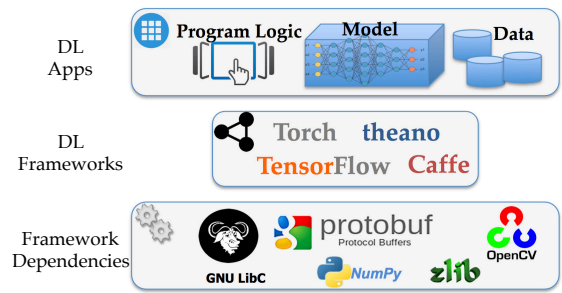

Fig. 1: The Layered Approach for Deep Learning Applications.

The exact implementation of deep learning applications varies, but those built on deep learning frameworks are usually consisted of software in three layers. Figure 1 shows the layers of typical deep learning applications. The top layer contains the application logic, the deep learning model and corresponding data resulted from the training stage. These are components usually visible to the developers. The middle layer is the implementation of the deep learning frameworks, such as tensor components and various filters. The interface between the top two layers are usually specified in the programming language used to implement the middle layer. For examples, the choices of programming language interfaces include C++, Python, and Lua for Caffe, TensorFlow, and Torch respectively. The bottom layers are building blocks used by the frameworks. These build blocks are components to accomplish tasks such as video and audio processing and model representations (e.g. protobuf). The selection of building blocks varies depending on the design of a framework. For example, TensorFlow contains its own implementations of video and image processing built over 3rd party packages such as librosa and numpy, whereas Caffe chooses to directly use open source libraries, such as OpenCV and Libjasper, to parse media inputs. Even the bottom and the middle layers are often invisible to the developers of the deep learning applications, these components are essential part of deep learning applications.

Table I provides some basic statistics of the implementations of deep learning frameworks. In our study, the versions of TensorFlow and Caffe that we analyzed are 1.2.1 and 1.0. The study also include Torch7. As the default Torch package only support limited image formats, we choose to study the version of Torch7 that combines OpenCV [9] that support various image formats such as bmp, gif, and tiff.

We measure the complexity of a deep learning framework by two metrics, the lines of code and the number of software dependency packages. We count the lines of code by using the *cloc* tool on Linux. As described in table I, all these implementation's code bases are not small. Tensorflow has more 887 thousands lines of code, Torch has more than 590K lines of code, and Caffe has more than 127K. In addition, they all depends on numerous 3rd party packages. Caffe is based on more than 130 depending libraries (measured by the Linux *ldd* utility), and Tensorflow and Torch depend on 97 Python modules and 48 Lua modules respectively, which was counted by the import or require modules.

TABLE I: DL frameworks and Their Dependencies

| DL Framework | lines of code | number of dep. package | sample packages |
|---|---|---|---|
| Tensorflow | 887K+ | 97 | librosa,numpy |
| Caffe | 127K+ | 137 | libprotobuf,libz,opencv |
| Torch | 590K+ | 48 | xlua,qtsvg,opencv |

Layered approach is a common practice for software engineering. Layering does not introduce flaws directly, but complexity in general increases the risks of vulnerabilities. Any flaw in the framework or its building components affects applications building on it. The next section of this paper presents some preliminary findings of flaws in implementations.

III. VULNERABILITIES AND THREATS

While there are numerous discussion about deep learning and artificial intelligence applications, the security of these applications draws less attention. To illustrate the risks and threats related to deep learning applications, we first present the attack surfaces of machine learning applications and then consider the type of risks resulted from implementation vulnerabilities.

*A. Attack Surfaces*

Without losing generality, here we use MNIST handwriting digits [11] recognition as an example to consider the attack surface of deep learning applications. We believe an image recognition application like MNIST can be exploited from the following three angles:

- Attack Surface 1 – *Malformed Input Image*: Many current deep learning applications, once being trained, usually work on input data for classification and recognition purposes. For an application that read inputs from files or the network, attackers potentially can construct malformed input. This applies to the MNIST image recognition application, which read inputs from files. The attack surface is significantly reduced for applications that take input from a sensor such as a directed connected camera. But the risk of malformed input is not eliminated in those cases, and we will discuss it in the next section.
- Attack Surface 2 – *Malformed Training Data*: Image recognition applications take training samples, which can be polluted or mislabeled if training data come from external sources. This is often known as data poisoning attack.
  Data poisoning attack does not need to rely on software vulnerabilities. However, flaws in implementations can make data poisoning easier (or at least harder to be detected). For example, we have observed inconsistency of the image parsing procedure in the framework and common desktop applications (such as image viewer). This inconsistency can enable a sneaky data pollution without being noticed by people managing the training process.
- Attack Surface 3 – *Malformed Models*: Deep learning applications can also be attacked if the developers use models developed by others. Although many developers design and build models from scratch, many models are made available for developers with less sophisticated machine learning knowledge to use. In such case, these models becomes potential sources that can be manipulated by attackers. Similar to data poisoning attacks, attackers can threat those applications carrying external models without exploiting any vulnerabilities. However, implementation flaws, such as a vulnerability in the model parsing code help attackers to hide malformed models and make the threat more realistic.

Certainly, the attack surface varies based on each specific application, but we believe these three attack surfaces cover most of the space from where attackers threat deep learning applications.

*B. Type of Threats*

We have studied several deep learning frameworks and found a dozen of implementation flaws. Table II summarizes a portion of these flaws that have been assigned with CVE numbers. These implementation flaws make applications vulnerable to a wide range of threats. Due to the space limitation, here we only present the threats caused by malformed input, and we assume the applications take input from files or networks.

TABLE II: CVEs Found for DL frameworks and Dependencies

| DL Framework | dep. packages | CVE-ID | Potential Threats |
|---|---|---|---|
| Tensorflow | numpy | CVE-2017-12852 | DOS |
| Tensorflow | wave.py | CVE-2017-14144 | DOS |
| Caffe | libjasper | CVE-2017-9782 | heap overflow |
| Caffe | openEXR | CVE-2017-12596 | crash |
| Caffe/Torch | opencv | CVE-2017-12597 | heap overflow |
| Caffe/Torch | opencv | CVE-2017-12598 | crash |
| Caffe/Torch | opencv | CVE-2017-12599 | crash |
| Caffe/Torch | opencv | CVE-2017-12600 | DOS |
| Caffe/Torch | opencv | CVE-2017-12601 | crash |
| Caffe/Torch | opencv | CVE-2017-12602 | DOS |
| Caffe/Torch | opencv | CVE-2017-12603 | crash |
| Caffe/Torch | opencv | CVE-2017-12604 | crash |
| Caffe/Torch | opencv | CVE-2017-12605 | crash |
| Caffe/Torch | opencv | CVE-2017-12606 | crash |
| Caffe/Torch | opencv | CVE-2017-14136 | integer overflow |

- Threat 1 – *DoS attacks* : The most common vulnerabilities that we found in deep learning frameworks are software bugs that cause programs to crash, or enter an infinite loop, or exhaust all of memory. The direct threat caused by such bugs are denial-of-service attacks to applications running on top of the framework. The list below shows the patch to a bug found in the *numpy* python package, which is a building block for the TensorFlow framework. The *numpy* package is used for matrix multiplication and related processing. It is commonly used by applications built over TensorFlow. The particular bug occurs in the *pad()* function, which contains a *while* loop that would not terminate for inputs not anticipated by the developers. The flaws occur because of the variable *safe-pad* in the loop condition is set to a negative value when an empty vector is passed from a caller. Because of this bug, we showed that popular sample TensoFlow applications, such as the Urban Sound Classification [7], will hang with special crafted sound files.

Listing 1: numpy patch example
```
--- a/numpy/lib/arraypad.py
+++ b/numpy/lib/arraypad.py
@@ -1406,7 +1406,10 @@ def pad(array, pad_width, mode, **kwargs):
             newmat = _append_min(newmat, pad_after, chunk_after, axis)

    elif mode == 'reflect':
-     for axis, (pad_before, pad_after) in enumerate(pad_width):
+     if narray.size == 0:
+        raise ValueError("There aren't any elements to reflect in 'array'!")
+
+     for axis, (pad_before, pad_after) in enumerate(pad_width):
           ... ...
           method = kwargs['reflect_type']
           safe_pad = newmat.shape[axis] - 1
           while ((pad_before > safe_pad) or (pad_after > safe_pad)):
               ... ...
```

- Threat 2 – *Evasion attacks*: Evasion attacks occur when an attacker can construct inputs that should be classified as one category but being misclassified by deep learning applications as a different category. Machine learning researchers have spent a considerable amount of research effort on generating evasion input through adversarial learning methods [5, 10]. When faced with vulnerable deep learning framework, attackers can instead achieve the goal of evasion by exploiting software bugs. We found multiple memory corruption bugs in deep learning frameworks that can potentially cause applications to generate wrong classification outputs. Attackers can achieve evasion by exploiting these bugs in two way: 1) overwriting classification results through vulnerabilities that given attackers ability to modify specific memory content, 2) hijacking the application control flow to skip or reorder model execution.

The list below shows an out-of-bounds write vulnerability and the corresponding patch. The *data* pointer could be set to any value in the *readData* function, and then a specified data could be written to the address pointed by *data*. So it can potentially overwrite classification results.

Listing 2: OpenCV patch example
```
bool BmpDecoder::readData( Mat& img )
{
    uchar* data = img.ptr();
    ....
    if( m_origin &=& IPL_ORIGIN_BL )
    {
        data += (m_height - 1)*(size_t)step; // result an out bound write
        step = -step;
    }
    ....
    if( color )
        WRITE_PIX( data, clr[t] );
    else
        *data = gray_clr[t];
    ....
}
index 3b23662..5ee4ca3 100644
--- a/modules/imgcodecs/src/loadsave.cpp
+++ b/modules/imgcodecs/src/loadsave.cpp
+
+static Size validateInputImageSize(const Size& size)
+{
+    CV_Assert(size.width > 0);
+    CV_Assert(size.width <= CV_IO_MAX_IMAGE_WIDTH);
+    CV_Assert(size.height > 0);
+    CV_Assert(size.height <= CV_IO_MAX_IMAGE_HEIGHT);
+    uint64 pixels = (uint64)size.width * (uint64)size.height;
+    CV_Assert(pixels <= CV_IO_MAX_IMAGE_PIXELS);
+    return size;
+}

@@ -408,14 +426,26 @@ imread_( const String& filename, int flags, int hdrtype, Mat* mat=0 )
    // established the required input image size
-    CvSize size;
-    size.width = decoder->width();
-    size.height = decoder->height();
+    Size size = validateInputImageSize(Size(decoder->width(), decoder->height()));
```

- Threat 3 – *System Compromise*: For software bugs that allows an attacker to hijack control flow, attackers can potentially leverage the software bug and remotely compromise the system that hosts deep learning applications. This occurs when deep learning applications run as a cloud service to input feed from the network.

The list below shows a patch to a simple buffer overflow found in the OpenCV library. The OpenCV library is a computer vision library which designed for computational efficiency and with a strong focus on real-time applications. OpenCV supports the deep learning frameworks, such as TensorFlow, Torch/PyTorch and Caffe. The buffer overflow occurs in the *readHeader* function in grfmt_bmp.cpp. The variable *m_palatte* represents a buffer whose size is 256*4 bytes, however, the value of *clrused* is from an input image which can be set to an arbitrary value by attackers. Therefore, a malformed BMP image could result to buffer overflow from the *getBytes()* call. Through our investigation, this vulnerability provides the ability to make arbitrary memory writes and we have successfully forced sample programs (such as

cpp_classification [2] in Caffe) spawning a remote shell based on our crafted image input.

While doing this work, we found another group of researchers [8] that have also studied the vulnerabilities and impact of OpenCV on machine learning applications. Although their idea of exploring OpenCV for system compromise shares a similar goal with our effort, they did not find or release vulnerabilities that are confirmed by OpenCV developers [4]. In contrast, our findings have been confirmed by corresponding developers and many of them have been patched based on our suggestion. In addition, we have also developed proof-of-concept exploitation that has successfully demonstrated remote system compromise (by remotely gaining a shell) through the vulnerabilities found by us.

Listing 3: OpenCV patch example

```
index 86cacd3..257f97c 100644
--- a/modules/imgcodecs/src/grfmt_bmp.cpp
+++ b/modules/imgcodecs/src/grfmt_bmp.cpp
@@ -118,8 +118,9 @@ bool BmpDecoder::readHeader()

    if( m_bpp <= 8 )
    {
-       memset( m_palette, 0, sizeof(m_palette));
-       m_strm.getBytes( m_palette, (clrused == 0? 1<<m_bpp : clrused)*4 );
+       CV_Assert(clrused < 256);
+       memset(m_palette, 0, sizeof(m_palette));
+       m_strm.getBytes(m_palette, (clrused == 0? 1<<m_bpp : clrused)*4 );
        iscolor = IsColorPalette( m_palette, m_bpp );
    }
    else if( m_bpp == 16 && m_rle_code == BMP_BITFIELDS )
```

## IV. DISCUSSION AND FUTURE WORK

The previous section presents software vulnerabilities in the implementations of deep learning frameworks. These vulnerabilities are only a set of factors that affect the overall application security. There are multiple other factors to consider, such as where does an application take input from, whether training data are well formatted, that also affect the security risks. We briefly discussed a few related issues here.

### A. Security Risks for Applications in Closed Environments

Many sample deep learning applications are designed to be used in a closed environment, in which the application acquires input directly from sensors closely coupled with the application. For example, the machine learning implementation running on a camera only takes data output from the built-in camera sensor. Arguably the risk of malformed input is lower than an application takes input from network or files controlled by users. However, a closely coupled sensor does not eliminate threats of malformed input. For example, there are risks associated with sensor integrity, which can be compromised. If the sensor communicates with a cloud server where the deep learning applications run, attackers could reverse the communication protocol and directly attack the backend.

### B. Detect Vulnerabilities in Deep Learning Applications

We applied traditional bug finding methods, especially fuzzing, to find the software vulnerabilities presented in this paper. We expect all conventional static and dynamic analysis methods apply to the deep learning framework implementation. However, we found that coverage-based fuzzing tools are not ideal for testing deep learning applications, especially for discovering errors in the execution of models. Taking the MNIST image classifier as an example, almost all images cover the same execution path because all inputs go through the same layers of calculation. Therefore, simple errors such as divide-by-zero would not be easily found by coverage-based fuzzers since the path coverage feedback is less effective in this case.

### C. Security Risks due to Logical Errors or Data Manipulation

Our preliminary work focused on the "conventional" software vulnerabilities that lead to program crash, control flow hijacking or denial-of-service. It is interesting to consider if there are types of bugs specific to deep learning and need special detection methods. Evasion attack or data poisoning attack do not have to relies on conventional software flaws such as buffer overflow. It is enough to create an evasion if there are mistakes allowing training or classification to use more data than what an application suppose to have. The mismatch of data consumption can be caused by a small inconsistency in data parsing between the framework implementation and the conventional desktop software.

One additional challenge for detecting logical errors in deep learning applications is the difficulty to differentiate insufficient training from intended manipulation, which targets to have a particular group of inputs misclassified. We plan to investigate methods to detect such type of errors.

## V. CONCLUSION

The purpose of this work is to raise awareness of the security threats caused by software implementation mistakes. Deep Learning Frameworks are complex software and thus it is almost unavoidable for them to contain implementation bugs. This paper presents an overview of the implementation vulnerabilities and the corresponding risks in popular deep learning frameworks. We discovered multiple vulnerabilities in popular deep learning frameworks and libraries they use. The types of potential risks include denial-of-service, evasion of detection, and system compromise. Although closed applications are less risky in terms of their control of the input, they are not completely immune to these attacks. Considering the opaque nature of deep learning applications which buries the implicit logic in its training data, the security risks caused by implementation flaws can be difficult to detect. We hope our preliminary results in this paper can remind researchers to not forget conventional threats and actively look for ways to detect flaws in the software implementations of deep learning applications.

## REFERENCES

[1] Gardener and Benoitsteiner, "An open-source software library for Machine Intelligence," https://www.tensorflow.org/, 2017.
[2] Y. Jia, "Classifying ImageNet: using the C++ API," https://github.com/BVLC/caffe/tree/master/examples/cpp_classification, 2017.
[3] Y. Jia, E. Shelhamer, J. Donahue, S. Karayev, J. Long, R. Girshick, S. Guadarrama, and T. Darrell, "Caffe: Convolutional architecture for fast feature embedding," *arXiv preprint arXiv:1408.5093*, 2014.
[4] Opencv Developers, "Opencv issue 5956," https://github.com/opencv/opencv/issues/5956, 2017, accessed 2017-09-03.
[5] N. Papernot, P. McDaniel, I. Goodfellow, S. Jha, Z. B. Celik, and A. Swami, "Practical black-box attacks against machine learning," in *Proceedings of the 2017 ACM on Asia Conference on Computer and Communications Security*, ser. ASIA CCS '17. New York, NY, USA: ACM, 2017, pp. 506–519.
[6] Ronan, Clément, Koray, and Soumith, "Torch: A SCIENTIFIC COMPUTING FRAMEWORK FOR LUAJIT," http://torch.ch/, 2017.
[7] A. Saeed, "Urban Sound Classification," https://devhub.io/zh/repos/aqibsaeed-Urban-Sound-Classification, 2017.

[8] R. Stevens, O. Suciu, A. Ruef, S. Hong, M. W. Hicks, and T. Dumitras, "Summoning demons: The pursuit of exploitable bugs in machine learning," *CoRR*, vol. abs/1701.04739, 2017. [Online]. Available: http://arxiv.org/abs/1701.04739

[9] VisionLabs, "OpenCV bindings for LuaJIT+Torch," https://github.com/VisionLabs/torch-opencv, 2017.

[10] W. Xu, Y. Qi, and D. Evans, "Automatically evading classifiers," in *Network and Distributed System Security Symposium*, 2016.

[11] L. Yann, C. Corinna, and J. B. Christopher, "The MNIST Database of handwritten digits," http://yann.lecun.com/exdb/mnist/, 2017.